\documentclass{elsart}
\usepackage{amssymb,amsmath,multirow,epsf,epsfig,graphicx,color}
\usepackage{calrsfs}    
\begin{document}
\begin{frontmatter}
\title{Neutron$-^{19}$C scattering: emergence of universal properties in a finite range potential}
\author[IFT]{M. A. Shalchi}
\footnote{Email: shalchi@ift.unesp.br},
\author[IFT]{M.T. Yamashita},
\author[Ohio,Ohio2]{M. R. Hadizadeh},
\author[ITA]{T. Frederico} and
\author[IFT,ITA]{Lauro Tomio}
\address[IFT]{
Instituto de F\'\i sica Te\'orica, UNESP - Univ. Estadual Paulista, \\ CEP 01140-070, S\~ao Paulo, SP, Brazil.}
\address[Ohio]{Institute for Nuclear and Particle Physics and Department of Physics and Astronomy, 
Ohio University, Athens, OH 45701, USA.}
\address[Ohio2]{College of Science and Engineering, Central State University, \\ Wilberforce,
OH 45384, USA.}
\address[ITA]{Instituto Tecnol\'ogico de Aeron\'autica,
DCTA, \\ 12228-900, S\~ao Jos\'e dos Campos, Brazil.}
\date{\today}
\maketitle
\begin{abstract}  
The low-energy properties of the elastic $s-$wave scattering for the $n-^{19}$C are studied
near the critical condition for the occurrence of an excited Efimov state in $n-n-^{18}$C. It is
established to which extent the universal scaling laws, strictly valid in the zero-range limit,
survive when finite range potentials are considered.
By fixing the two-neutrons separation energy in $^{20}$C with available experimental data, it is
studied the scaling of the real ($\delta_0^R$) and imaginary parts of the $s-$wave phase-shift
with the variation of the $n-^{18}$C binding energy. We obtain some universal characteristics
given by the pole-position of $k\cot(\delta_0^R)$ and effective-range parameters.
By increasing the $n-^{18}$C binding energy, it was verified that the excited state of $^{20}$C goes
to a virtual state, resembling the neutron-deuteron behavior in the triton.
It is confirmed that the analytical structure of the unitary cut is not
affected by the range of the potential or mass asymmetry of the three-body system. 
\end{abstract}
\end{frontmatter}

\section{Introduction}
\label{introduction}

The investigation of the low-energy continuum states of two neutrons and a core 
($n-n-c$) nucleus, which forms a weakly-bound state near or at the neutron drip line, can shed 
light on the role of the universal scaling laws~\cite{FrePPNP12} obeyed by  scattering observables, 
when the three-body system is close to forming an excited Efimov state~\cite{Efimov}.
The structure of halo nuclei turns out to be a fascinating topic, with impressive progress in recent
years, as one can trace from the topical review \cite{riisager2013}, where it is also discussed possible 
universal characterizations of such exotic nuclei.
In this respect, one may find relevant to move from negative energy states to the low energy scattering 
sector in order to evidence universal effects associated to the Efimov physics in the continuum. 

As an appropriate example of a system in which one can study universal properties, model independent 
ones,  in the scattering region, we can mention the $^{20}$C 
nucleus~\cite{YamNPA04}. This neutron-rich exotic nucleus described as $n-n-c$ system 
where  $n-^{18}$C forms a $s-$wave bound state, is close to having an Efimov excited state, 
as suggested in Ref.~\cite{FedorovPRL94}.
The possibility of a $s-$wave excited halo state in this nucleus was analyzed by means of the critical 
conditions proposed in Ref.~\cite{Amorim97}, which was further discussed in Refs.~\cite{hammer08,FrePPNP12}.
Within actual experimental perspectives, it was reported in ~\cite{MaccFBS15} that beams of $^{19}$C near barrier 
energies will be available at the FRIB facility~\cite{FRIB-MSU} with reaccelerated beam rates expected in the 
range 100$-$106 pps. This allows to experimentally explore the reaction $^{19}$C(d, p)$^{20}$C~\cite{MaccFBS15}
 (see the proposal ~\cite{MaccLOI14}), which we suggest can also
be appropriate to investigate the continuum state energies of $n-^{19}$C and $n-n-^{18}$C 
close to their respective thresholds. 

As in the present work we are concerned with the $n-^{19}$C elastic scattering properties within a  three-body 
framework, we found convenient to remind some original studies considering the neutron-deuteron 
($n-d$) system. In the study of the spin-doublet $s-$wave $n-d$ elastic scattering, a relevant quantity that  
was considered is $k\cot\delta_0$, where $k$ is the on-energy shell relative momentum 
(incoming and final) and $\delta_0$ is the $s$-wave phase shift. It was first pointed out
in Ref.~\cite{vanoers} that this quantity presents a pole when analyzing experimental
data for the $n-d$ doublet $s-$wave phase shift, motivating some other pioneering 
related works~\cite{reiner,wfuda,gfuda}.
In order to study the analytical structure of this quantity, Van Oers and Seagrave 
proposed to incorporate the pole in a phenomenological effective range formula,
by fitting $k\cot\delta_0$ to low energy data for the $n-d$ $s-$wave doublet state just 
below the elastic threshold, given by~\cite{vanoers}
\begin{equation}
 k\cot\delta_0 = -A + B k^2 - \frac{C}{1+D k^2}, \label{kcotnd}
\end{equation}
where $A$, $B$, $C$, and $D$ are fitted constants.  
The corresponding pole and residue have been obtained from dispersion relations 
as well as from exact solutions of three-particle equations with separable interactions by Whiting 
and Fuda~\cite{wfuda}. 

The existence of the triton virtual state was found on the basis of 
the effective range expansion~\cite{gfuda}. Later on in ~\cite{adhprc821}, within a three-body model with
finite range potentials, it was shown that the triton virtual state appears from an 
excited Efimov state moving to the non-physical energy sheet through the elastic 
cut when the deuteron binding increases. The same behavior was found in the case of the excited Efimov state in $^{20}$C, 
considering now the variation of the $^{19}$C binding energy and the associated $n-^{19}$C elastic cut~\cite{c20}. 
The excited state of $^{20}$C
appears either as a bound or virtual state, depending on the low-energy parameters 
of the two-body subsystems as well as a three-body scale, such as the ground state energy of
$^{20}$C.

The understanding of low-energy properties of $n-n-^{18}$C system also requires studying the scattering amplitude when an Efimov state is near the physical region. 
A relevant task is to find out the possible restriction on the expected general validity of  the effective-range expansion (\ref{kcotnd}) when considering mass-imbalanced three-particle
systems and finite-range potentials. 
The expected universal behavior of the elastic $s-$wave scattering amplitudes was also discussed for the atom-dimer system~\cite{braaten}, where the atom-dimer scattering length 
$a_{AD}$ can vanish or diverges when the shallow state moves by changing the atom-atom scattering length and/or the three-body parameter.

The results of the zero-range calculations given in  \cite{YamPLB08b}
for the $n-^{19}$C elastic $s-$wave phase-shifts support the existence of 
a pole in the effective range expansion in a good qualitative agreement with 
what is found for the $n-d$ low-energy scattering, as well as for the atom-dimer discussed
above. It remains the question how much such universal features are preserved for
finite-range potentials, considering the range of the $n-c$ and 
$n-n$ interactions, when the two-neutron separation energy in $^{20}$C is kept fixed.  

In the present work, in order to further understand the low-energy scattering behavior of the $s-$wave $n-^{19}$C scattering
(considered in Refs.~\cite{ArPRC04,c20}), we are considering a deeper investigation on this system, which goes beyond the zero-range approach that was used in Ref.~\cite{c20}, by 
considering a finite-range potential, which was chosen separable with Yamaguchi form.
Finite-range effects for the spectrum of three-boson systems have been studied by several groups, considering potential models near the unitary limit. For instance, in Ref.~\cite{hammer08} the authors have considered range effects in the zero-range approach presented in Ref.~\cite{Amorim97}. Next, we can also mention the studies
presented in Ref.~\cite{jensen} on the conditions for Efimov physics when using finite-range potentials.
In Refs.~\cite{KieFBS11,KiePRA15}, and more recently in ~\cite{KiePRA16}, one can find results of several investigations
on the universal behaviors of the spectrum of up to four-boson systems by using potential models.
Our work  complement such studies by investigating the consequences for the s-wave scattering amplitude associated with 
the movement of an excited Efimov state towards the continuum as  the bound/virtual energy  of the two-body subsystem 
is changed with respect to the three-body bound state energy.
Another relevant task in the present approach is to study the effect of mass asymmetry associated with finite-range interactions in 
the behavior of the low-energy pole $ k\cot\delta_0^{R}$, which is a quantity well characterized for the  $n-d$ system, as discussed before.
We will address the scaling of the real $(\delta_0^R)$ and the imaginary parts of the phase-shift with the variation of the $^{19}$C halo 
neutron binding energy when the two neutrons separation energy in $^{20}$C is fixed to its experimental value. The scaling of the position 
of the pole in $k\cot{\delta_0^R}$ and the effective range parameters reproduces universal characteristics found the zero-range model under 
the realm of the Efimov physics. We also investigate how the Efimov excited state of $^{20}$C moves to a virtual state by increasing the 
$^{19}$C halo neutron binding energy, as found for the zero-range model by the introduction of a finite range potential. We remind that this 
was also studied before in Refs.~\cite{adhprc821} for the case of triton within a separable model.

In the next section, we present the basic formalism for a three-body halo nucleus compounded by a core and two neutrons interacting via two-body separable potential. The main results are presented in 
Sect.~\ref{results}, followed by our conclusions in Sect.~\ref{conclusions}.

\section{Formalism and notation}
\label{formalism}
In the present section, for the sake of completeness and to fix our notation, we include the standard 
formalism for the elastic scattering amplitude of a neutron in the bound neutron-core subsystem. 
For the input in the scattering equations we have the $n-n$ virtual-state energy ($E_{nn}= -$143 keV), the $n-c$ subsystem bound-state energy ($E_{nc}=E_{^{19}{\rm C}}$), and the $n-n-c$ three-body
ground-state binding energy ($E_{^{20}{\rm C}}=-$ 3.5 MeV).
The neutron separation energy in $^{19}$C has a sizable error, with given values ranging from $-$160$\pm$110 keV~\cite{audi} to $-$530$\pm$130 keV~\cite{naka99}. Therefore, in our present study, a wide variation for the $n-^{18}$C bound-state energy will be allowed, within a range between 200 up to 850 keV, such that all possible experimental values can be included.

Our units are such that $\hbar=1$, with momentum variables in fm$^{-1}$ and the unit conversion given by $\hbar^2/m_n= 
41.47$ MeV fm$^2$. In the formalism, the core-mass number is defined as $A=m_c/m_n$, with $\mu_{nc}=Am_n/(A+1)$ being the reduced mass for the $n-c$ system. Therefore, in our calculations, for 
the specific system we are considering, $A=18$ and $\mu_{nc}=(18/19)m_n$.
The energies for the two-body subsystems are given by $E_{nn}$ (virtual) and $E_{nc}$ (considered bound), 
with the total three-body energy given by $E_3\equiv E$.
The $s-$wave elastic $n-(nc)$ scattering formalism, discussed in detail in Ref.~\cite{c20} for a zero-range potential, is extended to include finite-range two-body interactions, assumed having the following rank-one separable Yamaguchi format:
\begin{eqnarray}
 V_{ij}(p,p')=\lambda_{ij}\left(\frac{1}{p^2+\beta_{ij}^2}\right)\left(\frac{1}{p'^2+\beta_{ij}^2}\right),
\end{eqnarray}
where $ij=nn$ or $nc$, respectively, for the $n-n$ or $n-c$ two-body subsystems.
$\lambda_{ij}$ and $\beta_{ij}$ refer to the strength and range $r_{ij}$ of the respective two-body interaction.  
By considering negative two-body energies (bound or virtual), $E_{ij}$, the corresponding relations for the strengths and ranges are 
\begin{eqnarray}
 \lambda^{-1}_{ij}=\frac{-2\pi\mu_{ij}}{\beta_{ij}(\beta_{ij}\pm\kappa_{ij})^2},~~~~~~
 r_{ij}=\frac{1}{\beta_{ij}}+\frac{2\beta_{ij}}{(\beta_{ij}\pm\kappa_{ij})^2},
\end{eqnarray}
where $\kappa_{ij}=\sqrt{-2\mu_{ij} E_{ij}}$, with $(+)$ for bound and $(-)$ for virtual states, with $\mu_{ij}$ the reduced two-body mass.

The spectator function $\chi_n(\vec q)$, which represents the relative motion between the neutron and the $n-c$ subsystem target, within the elastic scattering boundary condition, is given by 
\begin{equation}
\chi_n(\vec q)\equiv(2\pi)^3\delta(\vec q-\vec k_i)
+4\pi\frac{h_n(\vec q;E(k_i))}{q^2-k_i^2-{\rm i}\epsilon} ,
\end{equation}
where $h_n(\vec q;E(k_i))$ is the scattering amplitude, and 
$\vec{q}$ is the momentum of the spectator particle ($n$) with respect 
to the center-of-mass (CM) of the other two particles ($n-c$).
The on-energy-shell incoming and final relative momentum are
related to the three-body energy $E \equiv E(k_i)$ by
$k\equiv k_i\equiv |\vec k_i|=|\vec k_f| =  \sqrt{2\mu_{n(nc)}\left(E-E_{nc}\right)}$, where $\mu_{n(nc)}=m_n(A+1)/(A+2)$. 
The coupled $s-$wave scattering equations can be cast in a
single channel Lippmann-Schwinger-type equation for the relevant amplitude $h_n$, with the following form:
\begin{eqnarray}
h_n(q,E)&=&  \mathcal{V} \,(q,k,E)+\frac{2}{\pi}\int_0^\infty dq' \, q'^2  \, \frac{\mathcal{V} \,(q,q',E)h_n(q',E)}{q'^2-k^2-{\rm i}
\epsilon},\end{eqnarray}
where the kernel of this integral equation for the  $n-(nc)$ channel amplitude contains the contribution of the coupled $(nn)-c$ channel.
This is given by
\begin{equation}
\label{V-equation}
\mathcal{V}(q,q',E)=\frac{\pi}{2}\bar{\tau}_{nc}(q)\left [K_2 \,(q,q',E)+\hspace{-0.2cm}\int_0^\infty dk \, k^2  \, K_1 \,(q,k,E)\tau_{nn}(k)K_1 \,(q',k,E)\right], 
\end{equation}
where we use the following definitions:
\begin{eqnarray}
\bar{\tau}_{nc}(q)&=&\frac{-\beta_{nc}(\beta_{nc}+\kappa_{nc})^2(\beta_{nc}+\kappa_{3nc})^2(\kappa_{nc}+\kappa_{3nc})(A+1)^2}{\mu_{nc}\pi(2\beta_{nc}+\kappa_{3nc}+\kappa_{nc})A(A+2)}, \cr
 \tau_{nn}(q)&=&\frac{2\beta_{nn}}{\mu_{nn}\pi}\frac{(\beta_{nn}+\kappa_{nn})^2(\beta_{nn}+\kappa_{3nn})^2}{(-2\beta_{nn}-\kappa_{3nn}+\kappa_{nn})(\kappa_{nn}+\kappa_{3nn})},  
\end{eqnarray}
with
\begin{eqnarray}
   \kappa_{nn}&=&\sqrt{-m E_{nn}},\;\;\; \kappa_{nc}=\sqrt{-\frac{2mA}{A+1}E_{nc}}\cr
   \kappa_{3nn}&=&\sqrt{-m(E-\frac{(A+2)q^2}{4Am})},\;\;\; \kappa_{3nc}=\sqrt{-\frac{2mA}{A+1}(E-\frac{(A+2)q^2}{2(A+1)m})},
\end{eqnarray}
and
\begin{eqnarray}
 K_1(q,q',E)&=&\int_{-1}^1 dx \,\frac{(q^2+\frac{q'^2}{4}+qq'x+\beta_{nn}^2)^{-1}(q'^2+\frac{q^2A^2}{(A+1)^2}+\frac{2qq'Ax}{(A+1)}+\beta_{nc}^2)^{-1}}{E+{\rm i}\epsilon-\frac{q^2}{m}-\frac{q'^2(A+1)}{2Am}-\frac{qq'x}{m}},\nonumber \\ \\
 K_2(q,q',E)&=&\int_{-1}^1 dx \,\frac{(q'^2+\frac{q^2}{(A+1)^2}+\frac{2qq'x}{(A+1)}+\beta_{nc}^2)^{-1}(q^2+\frac{q'^2}{(A+1)^2}+\frac{2qq'x}{(A+1)}+\beta_{nc}^2)^{-1}}{E+{\rm i}\epsilon-\frac{q^2(A+1)}{2Am}-\frac{q'^2(A+1)}{2Am}-\frac{qq'x}{Am}}\nonumber 
.\end{eqnarray}
For three-body system we have also a coupled equation, which can be written as a single one for both bound and virtual states($I=b,v$):
\begin{eqnarray}
h_{nc}(q)&=& 2 k_v \mathcal{V} \,(q,-ik_v,E_{3v})h_{nc}(-ik_v)\delta_{I,v}+\frac{2}{\pi}\int_0^\infty dq' \, q'^2  \, \mathcal{V} \,(q,q',E)\frac{h_{nc}(q')}{q'^2+k_I^2} \nonumber \\
\end{eqnarray}

In our calculations, we have adjusted the  parameters of the neutron-neutron separable interaction by fixing the virtual state energy to -143 keV, and by considering two values of the effective range: $r_{nn}=0.123$ fm, which is obtained with  $\beta_{nn}=24.5 fm^{-1}$;
 and $r_{nn}=2.37$ fm resulting  from $\beta_{nn}=1.34 fm^{-1}$. 
For each value of $E_{nc}\equiv E_{^ {19}C}$, we change $\beta_{nc}$ in order to keep the two neutron separation energy
of $^ {20}$C at 3.5 MeV\cite{audi}. In this way, we are able to have two solutions for each $E_{nc}$ for high and low values of $\beta_{nc}$, allowing the change in the neutron-core interaction range to large values close to the $^{18}$C  matter root-mean-square radius of 2.82$\pm$0.04 fm \cite{OzaNPA01}.

The numerical method to calculate the elastic scattering amplitude
relies on the use of an auxiliary function~\cite{adhprc81}, which is
a solution of an auxiliary integral equation similar to the original one, but with a kernel without the unitarity cut, due to a subtraction procedure at the momentum $k$.  
The scattering amplitude is obtained by evaluating the corresponding integrals over the auxiliary function. In this way, the two-body unitarity cut appears no more inside the integral equations, but 
just in simple integrals. In the present case, we have the following integral equation for the auxiliary function $\Gamma$, with the corresponding solution for
$h_n(q;E)$:
\begin{eqnarray}
{\Gamma_n}(q,k;{E})&=&{\mathcal{V}}(q,k;E)
+\frac{2}{\pi}\int_0^\infty\hspace{-0.1cm} dp \left[p^2{\mathcal
V}(q,p;E)- k^2{\mathcal{V}}(q,k;E)\right]
\frac{{\Gamma_n}(p,k;E)}{p^2-k^2},
\nonumber\\
{h_n}(q;E)&=&
\frac{{\Gamma_n}(q, k;E)}
{\displaystyle 1-\frac{2}{\pi}{k^2}\int_0^\infty dp
\;\frac{{\Gamma_n}(p,{ k};E)}{p^2-k^2-{\rm
i}\epsilon}} .\label{hna2}
\end{eqnarray}

For the on-shell scattering amplitude, we have
\begin{eqnarray}
&&h_n(k;E)= \frac{e^{{\rm i}\delta_0}\sin\delta_0}{k}=\frac{1}{k\cot\delta_0 - {\rm i}k},
\label{a3} 
 \end{eqnarray}
such that, from Eq.~(\ref{hna2}), we have
\begin{eqnarray}
&& k\cot\delta_0
= \frac{1}{{\Gamma_n}(k,k;E)} 
\left[1-
\frac{2}{\pi}k^2\int_0^\infty dp
\;\frac{{\Gamma_n}(p,k;E) - 
{\Gamma_n}(k,k;E)}{p^2-k^2}\right].
\end{eqnarray}
Note that the numerical stability and accuracy of the results are quite sensitive to the approach one is considering when an Efimov state is close to the scattering region. In this situation, the method used in Ref.~\cite{adhprc81}, as outlined above, is found far more accurate than the one by using contour deformation techniques.

\section{Results and discussion}
\label{results}

Our goal is to investigate the effect of the two-body interaction range on universal properties of unbalanced-mass 
three-body systems ($n-n-c$), which can be verified by considering the exotic halo-nuclei $^{20}$C.
This halo nucleus, within the $n-n-^{18}$C configuration, is known to be close to the critical condition to have an excited 
Efimov state (see e.g. \cite{FrePPNP12}), and therefore sensitive to the increase of the attractive interaction range. 
In order to analyze the possible limitation on the universal behavior, namely the departure of the zero-range model 
predictions for the excited Efimov bound or virtual state of $^{20}$C, 
we adopt the following systematic. The neutron-core interaction-range parameter, $\beta_{nc}$, of the one-term separable potential, is varied for different $E_{^{19}{\rm C}}$, by keeping fixed the low-energy observables $S_{2n}$ and the virtual state energy $E_{nn}$. For that, we study situations with two distinct parameter regions with
low and high values for  $\beta_{nn}$ and $\beta_{nc}$, which  correspond to larger and smaller effective ranges of the $n-n$ and $n-c$ interactions, respectively.

\begin{table}[thb!]
  \caption{One-neutron separation energy in $^{19}$C (left column), obtained by the given range parameters of the separable Yamaguchi potential, $\beta_{nc}$ (second and fourth columns), with the corresponding effective ranges (third and fifth columns).
The values of $\beta_{nc}$ were obtained by fitting the two-neutron separation energy in $^{20}$C (3.5 MeV~\cite{audi}), with the 
$n-n$ interactions fixed by the virtual state energy, $E_{nn}= -$143 keV.
 The respective fixed values of $\beta_{nn}$ for high and low ranges are 1.34 fm$^{-1}$ ($r_{nn}=2.372$ fm) and 
 24.50 fm$^{-1}$ ($r_{nn}=$0.1228 fm).
}\begin{center}
\begin{tabular}{c||cc||cc}
\hline\hline
& \multicolumn{2}{c||}{$\beta_{nn}=$1.34 fm$^{-1}$}
& \multicolumn{2}{c}{$\beta_{nn}=$24.5 fm$^{-1}$}
\\
$|E_{^{19}{\rm C}}| $(keV)& $\beta_{nc} ({\rm fm}^{-1})$  & $r_{nc}$(fm)& $\beta_{nc}({\rm fm}^{-1})$&$r_{nc}$(fm)   \\
\hline\hline
 200& 0.971 & 2.736 &18.970& 0.157\\
 400& 0.754 & 3.233 & 17.036& 0.174\\
 600&0.598  & 3.720 & 15.592& 0.190\\
 800& 0.477 & 4.255 &14.395& 0.205\\
\hline\hline
\end{tabular}
\end{center}
\label{table1}
 \end{table}

In Table~\ref{table1}, by keeping fixed the three-body bound-state energy, $E_{^{20}{\rm C}}=-$3.5 MeV, as well as the $n-n$ 
parameters, such that $E_{nn}=$-143 keV, we report a few values for the one-neutron separation energies in $^{19}$C, given by $E_{^{19}{\rm C}}$, with the respective values for the parameter $\beta_{nc}$ and corresponding effective ranges, given by $r_{nc}$. For the $n-n$ parameters, we have fixed $\beta_{nn}=1.34$ fm$^{-1}$ for the high-range cases, and $\beta_{nn}=24.50$ fm$^{-1}$ for the low-ranges ones.
For lower values of $\beta$'s, the effective range of the neutron-core interaction is comparable to the $^{18}$C   root-mean-square 
matter radius, $r_m[^{18}C]$= 2.82$\pm$0.04 fm \cite{OzaNPA01}, while for high values of $\beta$'s, $r_m[^{18}C]$ is quite large compared to 
the effective-range, such that it is not a realistic situation. The case when the $\beta$'s for both $n-c$ and $n-n$ interactions are large,
in the sense that the values of the corresponding effective ranges are much smaller than the core size and the $n-n$ effective range, 
$2.75$ fm \cite{MilPRP90}, the Yamaguchi separable model should approach the zero-range model, with respect to the low-energy 
properties of the $n-n-c$ system. We observe that by increasing the neutron binding in $^{19}$C $\beta_{nc}$ decreases and the effective 
range increases, while the two-neutron binding energy in $^{20}$C is fixed. This trend is understood by recalling the Thomas effect,
which says that the three-body binding in the zero angular momentum state decreases when the range increases, therefore the attraction 
behind this effect has to be softened if the two-body binding is increased, with a three-body binding fixed implying in an increase of the range. This qualitative behavior is clearly seen in Table~\ref{table1}.

\begin{figure}[h]
\begin{center}
\includegraphics[scale=0.33]{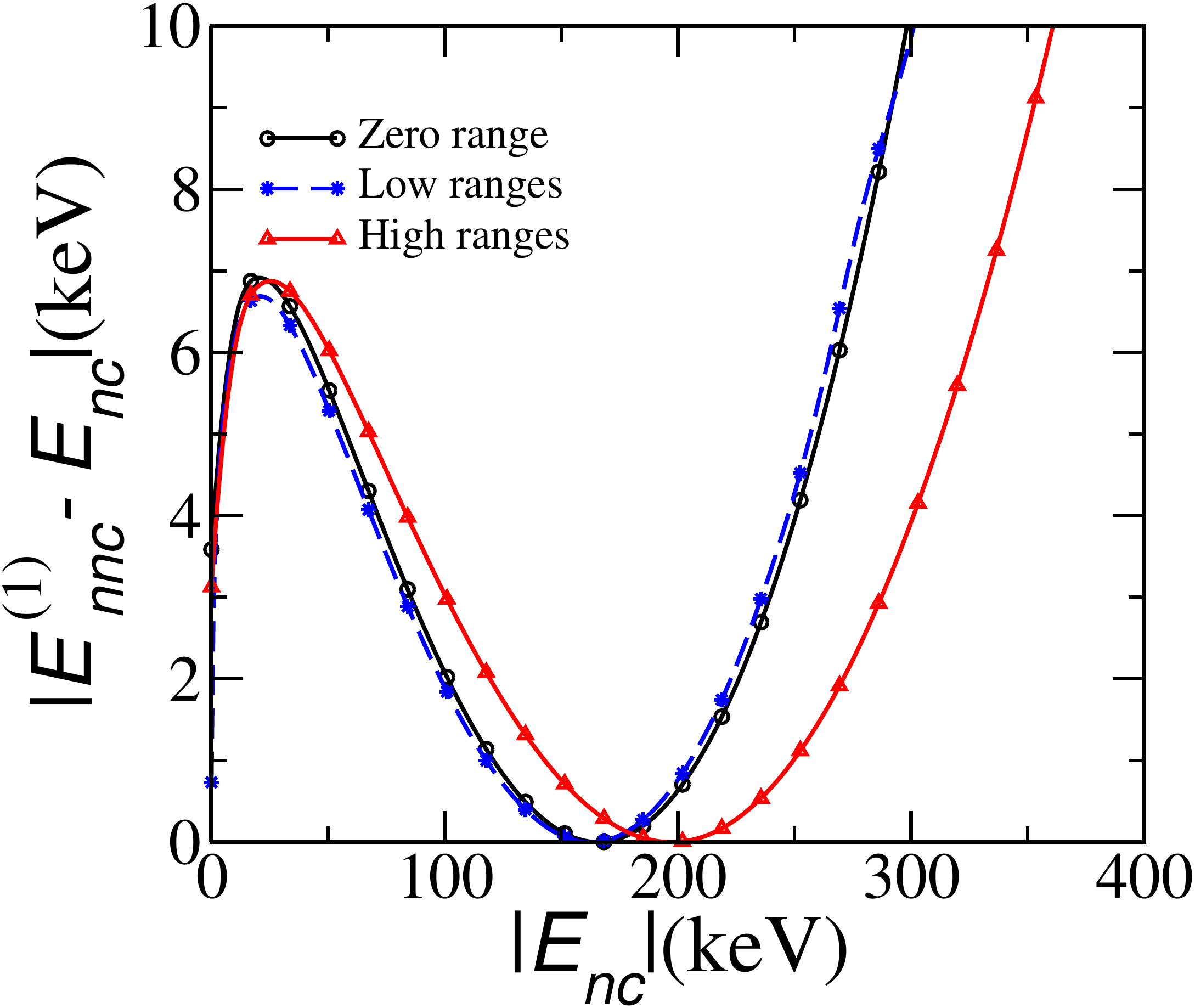}
\end{center}
\caption{Relation between the energy of the first excited (bound or virtual) state, $E^{(1)}_{^{20}{\rm C}}$, and the  $E_{n-^{18}{\rm C}}$ energies, indicating the threshold position when $E^{(1)}_{^{20}{\rm C}}$ = $E_{^{19}{\rm C}}$.
By increasing the two-body binding, $E_{^{19}{\rm C}}$ (decreasing $|a_{n-^{18}\rm{C}}|$), we reach the values for which 
an excited Efimov state $E^{(1)}_{^{20}{\rm C}}$ turns into a virtual state.  As shown, the corresponding position in the case
of zero-range results (black-solid lines with bullets), $|E_{nc}|=$ 167 keV, are close to the position obtained by the low-range (high-$\beta$) results (blue line with *). The high-range (low-$\beta$) results (red line with triangles) indicate that the threshold position is reached for higher values of  the $n-c$ binding, going from 167 to about 190 keV as the 
$n-c$ range is increased to about 2.7 fm.
} \label{fig:virtual}
\end{figure}

We start with the  calculation of the energy of the first excited state of $^{20}$C, by using the range parameters given in Table \ref{table1} in the separable Yamaguchi potentials for the $n-n$ and $n-c$ systems, in order to study  the effect of the interaction range in the general behavior of the excited state as the neutron-$^{18}$C binding energy is moved.  Close to the Efimov limit, the increase of the neutron-core binding energy for a fixed ground state $^{20}$C energy should make the excited state approach the continuum threshold and become a virtual state. We want to address to which extend this general trend is distorted, or even destroyed, with the increase of the interaction range.
The results  presented in Fig. \ref{fig:virtual} illustrate that. We compare the results with the universal zero-range model  obtained by calculations performed with renormalized subtracted equations,
explained in detail in Ref.~\cite{FrePPNP12}. For high-$\beta$ values (given in the  fourth column of Table \ref{table1}), when the 
$n-n$ and $n-c$ ranges are of the order 0.1 to 0.2 fm, the results are indeed quite compatible with the zero-range model ones, 
as expected. We should notice that the  $n-c$ energy required to have the excited state at the threshold, $|E_{nc}|=$ 167 keV, is moved to 190 keV when the $n-c$ range is increased to about 2.7 fm. By increasing the range, one naturally adds more attraction to the system and the excited state survives up to larger values of $|E_{nc}|$ for a fixed ground state energy, however the general picture provided by the scaling law obtained with the zero-range model is still valid, namely for an  $n-c$ effective range compatible with the matter radius of $^{18}$C.

Next, we  study the validity of the picture found with the zero-range model, therefore within the realm of Efimov physics
 for the elastic $s-$wave n-$^{19}$C scattering amplitude, when the interaction range is finite and compatible with the nuclear force. In order to perform this analysis we follow the suggestion given in Ref.~\cite{vanoers} presented in Eq. (\ref{kcotnd}), 
for the effective-range expansion of $k\cot\delta_0^R$ given in a re-parametrized form in \cite{YamPLB08b}, where $k$ is the on-energy-shell momentum for the $n-{^{19}C}$ scattering, and $\delta_0^R$ is the 
 real part of the corresponding $s-$wave phase shifts. $k\cot\delta_0^R$ is given in terms of the center of mass $n-{^{19}C}$ scattering energies $E_K\equiv k^2/(2\mu_{n,nc})$,  
  where $\mu_{n,nc}$ is the reduced 
  mass, $\mu_{n,nc}=m\,(A+1)/(A+2)$, which can be written as
\begin{equation}
k\cot\delta_0^{R}=\frac{-a^{-1} +b\; E_K  + c\;E_K^2}{1-E_K/E_0},
  \label{kcotd}
\end{equation}
where $a^{-1}$ is the two-body scattering length, $b$ and $c$ are the adjustable parameters, with $E_0$ giving the pole position.
The above effective-range expansion can also be expressed in a form close to the original expression (\ref{kcotnd}):
\begin{equation}
k\cot\delta_0^{R}=
\frac{d}{1-\frac{E_K}{E_0}}+e+f\,\frac{E_K}{E_0},
\label{kcotd2}
\end{equation}
where the residue is $d=-\frac{1}{a}+bE_0+cE_0^2$, the effective scattering length is $e= -bE_0-cE_0^2$ and $f=-cE_0^2$. 

\begin{figure}[tbh!]
\begin{center}
\includegraphics[scale=0.6]{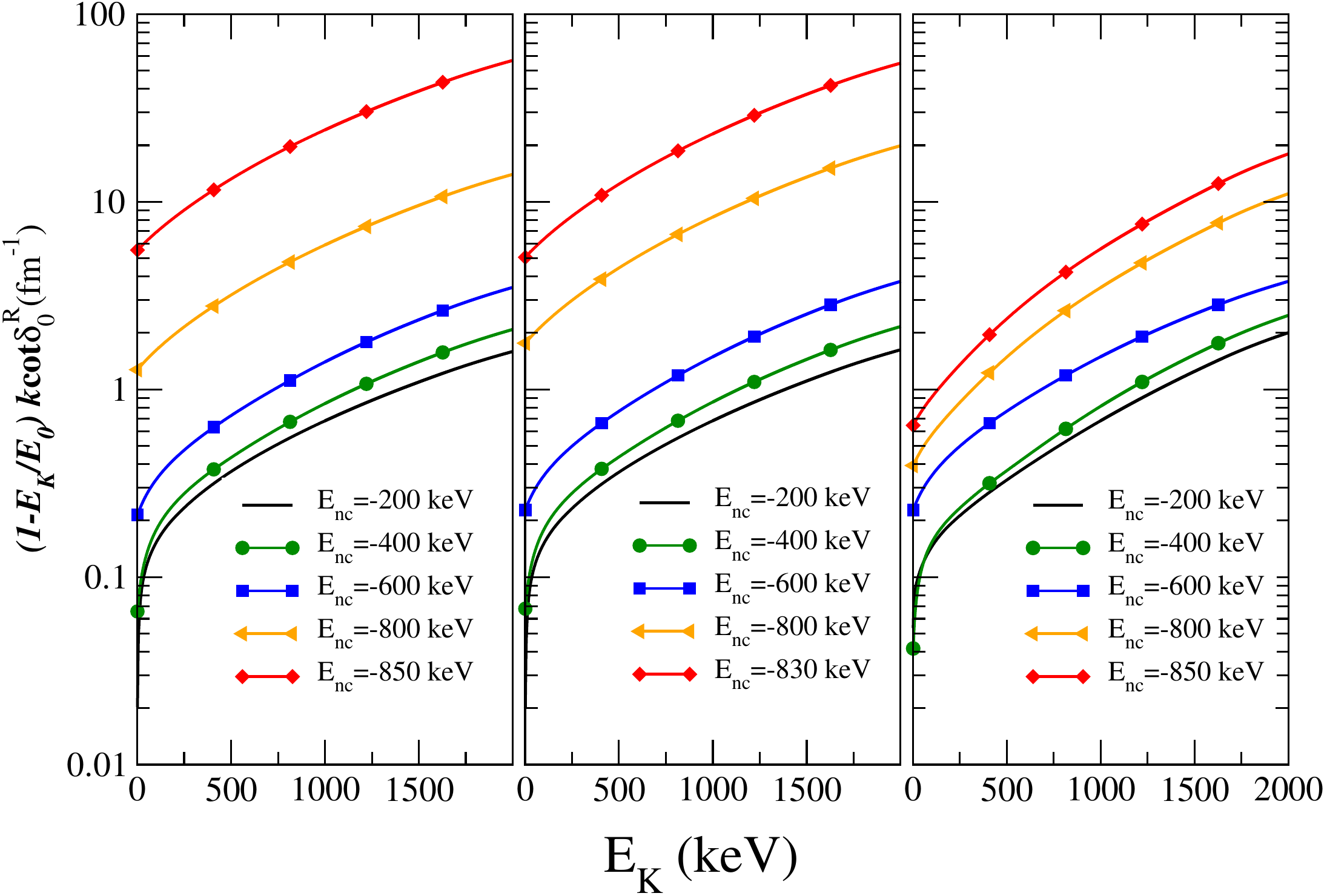}
\end{center}
\caption{
The function $(1-E_K/E_0) k\cot\delta_0^R$, where $E_0$ is the pole position, is given in terms of $E_K\equiv k^2/(2\mu_{n,nc})$, 
by considering renormalized zero-range potential (left frame), low-range Yamaguchi potential (middle frame, with large $\beta$ values), and high-range Yamaguchi potential (right frame, with small $\beta$ values.
For each case, we plot the results obtained for different values of the two-body binding energy $E_{nc}$, from -200 keV to
-850 keV, as shown inside the middle frame (the line conventions are the same for the three frames).
}
\label{figkcotze}
\end{figure}

The parameters of the expression (\ref{kcotd}) are obtained by fitting the results from the solution of the n-$^{19}$C scattering integral equation (\ref{hna2}) for the Yamaguchi one-term separable potential, obtained for different energies $E_K$. For that, we first present in Fig.~\ref{figkcotze} our results for the function $(1 - E_K/E_0) k \cot\delta^R_0$, from calculations performed for different $n-c$ energies, $E_{^{19}C}$, and fixed $E_{^{20}C}$ binding energy. 
The results shown in the figure are for the renormalized zero-range model \cite{YamPLB08b} (given for comparison) and for the Yamaguchi 
potential with high-$\beta$ and low-$\beta$ parametrizations given in Table~\ref{table1}.  
The function $(1 - E_K/E_0) k \cot\delta^R_0$ for Yamaguchi potential becomes similar to the one obtained with the zero-range model
when considering high values of $\beta$ (low-range). However, 
the values change when considering high-range parametrizations, with low values of $\beta$. 

The striking feature is the decrease of the sensitivity of $(1 - E_K/E_0) k \cot\delta^R_0$ on the different values of $E_{^{19}C}$.
While for the smallest binding energy of $^{19}$C namely 200 keV, the results from small-$\beta$'s are still quantitatively 
quite close to the ones calculated for large-$\beta$'s and zero-range model,  the differences  become larger as the $^{19}$C  
binding energy increases, but keeping the general trend. For small $|E_{^{20}C}|$ the Efimov physics should be certainly more dominant
explaining why the results are closer to the zero-range ones. The increase of the interaction range diminishes the region 
where the Efimov long range potential acts weakening the effect of such universal force in the three-body system. We remind that 
the action of such universal potential is present in region between the potential range up to the size of the $^{19}$C one-neutron halo $\sim \left[\sqrt{m|E_{^{19}C}|/\hbar^2}\right]^{-1}$, and thus the increase in the potential range weakens the effect of the universal potential and consequently the sensitivity to the variation of the $^{19}$C binding, 
which  weakens the dependence of $(1 - E_K/E_0) k \cot\delta^R_0$ with $|E_{^{19}C}|$ for a fixed $^{20}$C binding energy, as seen in Fig.~\ref{figkcotze}.

\begin{figure}[tbh!]
\begin{center}
\includegraphics[scale=0.33]{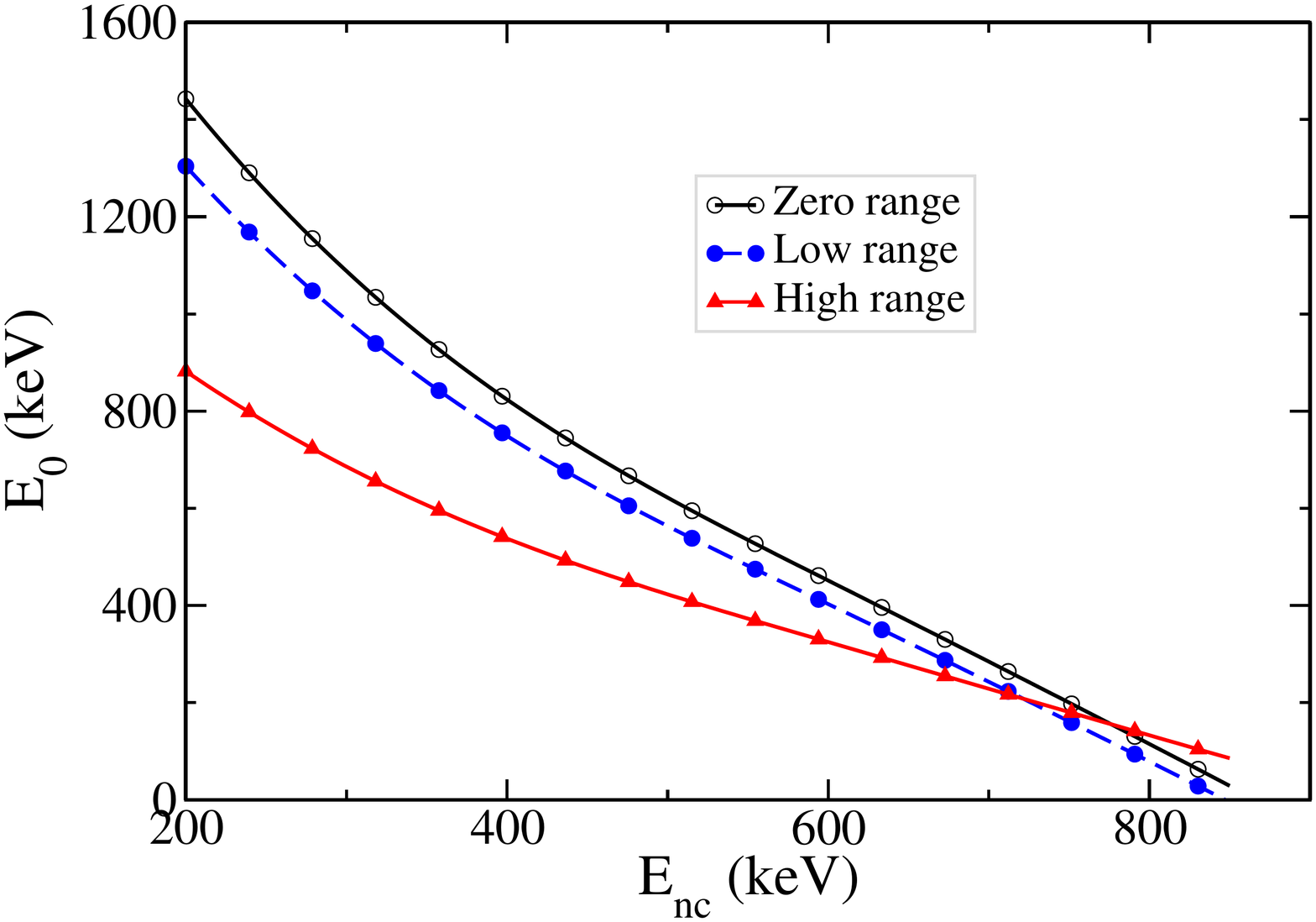}
\end{center}
\caption{Positions of the pole in $k\cot\delta_0^R$, given by $E_0$, shown as functions of the absolute values of the 
$n-^{18}$C binding energy for zero-range (black-solid line), low-range (high-$\beta$, blue-dashed line) 
and high-range (low-$\beta$, red-solid line) interactions.}
\label{figpoloc19c20}
\end{figure}

The position of the pole $E_0$ is plotted in Fig.~\ref{figpoloc19c20}, 
as a function of the corresponding $n-c$ binding energies, for the zero-range, high-beta (low-range) and low-beta (high-range) results. Although there is less sensitivity of $E_0$ with respect to $E_{^{19}C}$ for larger potential ranges in comparison to the low and zero-range results, still, the Efimov physics has a strong dominance on this observable. 
 It was suggested in \cite{YamPLB08b} to extend to the $n-n-c$ system the formula of the ratio found for three-boson system \cite{braaten},
 between the two-boson scattering lengths, $a_B$ where one Efimov state is at the threshold, and $a_0$ where the
atom-dimer scattering length is zero, or the pole in $k\cot\delta_0$ is at zero scattering energy, namely,
\begin{equation}
a_B/a_0=\exp\left(\frac{\pi/2-0.59654}{s_0}\right) .
\end{equation}
 For the $n-n-^{18}$C case, we considered $a_{nn}^{-1}=0$, with $s_0=1.12$ (in the case that $A=$ 18).
We should emphasize now that the quantities $a_B$ and $a_0$ will correspond to the $n-^{18}$C scattering 
lengths. In the limit of very large scattering lengths, the ratio $a_B/a_0$ is  
$\sqrt{E^0_{^{19}C}/E^B_{^{19}C}}$,
where $E^0_{^{19}C}$ is the value of the energy where the pole is at the scattering threshold and $E^B_{^{19}C}$ is the energy where the excited state of $^{20}$C is at the scattering threshold,
\begin{equation}
\sqrt{E^B_{^{19}C}/E^0_{^{19}C}}\approx\exp\left(-\frac{\pi/2-0.59654}{1.12}\right)=0.419 \, .\label{ratio}
\end{equation}
For the low-range potential, $E^0_{^{19}C}=850$ keV and $E^B_{^{19}C}=167$ keV resulting 
$\sqrt{E^B_{^{19}C}/E^0_{^{19}C}}=$ 0.44 and for high-range potential $E^0_{^{19}C}=940$ keV 
and $E^B_{^{19}C}=190$ keV resulting $\sqrt{E^B_{^{19}C}/E^0_{^{19}C}}=$ 0.45, and both cases the results compares
well with (\ref{ratio}).

In the Tables~\ref{table2} and \ref{table3} we present the extracted effective-range parameters, which are obtained by 
fitting the results shown in  Fig.~\ref{figkcotze} to Eq.~(\ref{kcotd}). The effect of the range can be observed by comparing the values
of both tables, where the high-$\beta$ values (low range) are given in Table~\ref{table2}, with the low-$\beta$  values (high range) 
given in Table~\ref{table3}. For reference, we also provide besides the position of the pole also the residue.
The pole in $ k \cot\delta^R_0$ moves below the breakup threshold as the $^{19}$C binding energy increases, and when 
$0<E_0<|E_{^{19}{\rm C}}|$ it should be observable as a zero in the $s-$wave cross-section at the center of mass energy $E_0$. 
That happens for $|E_{^{19}{\rm C}}|$ larger than  527 keV and  467 keV in  cases of the low- and high-ranges potentials, 
respectively. In both cases it happens close to the $E_{^{19}{\rm C}}$  experimental value of $-$530$\pm$130 keV from Ref. \cite{naka99}.
Furthermore, the Tables illustrate, the depletion of the  sensitivity  of the parameters in Eq. (\ref{kcotd}) with $E_{^{19}{\rm C}}$
for the $E_{^{20}{\rm C}}$ fixed when the range increases, as already discussed.

\begin{table}[thb!]
  \caption{Effective-range parameters, obtained by fitting Eq.~(\ref{kcotd}) to Fig.~\ref{figkcotze}, when considering different 
 values of $|E_{^{19}{\rm C}}|$ (first column) with  short-range  Yamaguchi potentials.}
\begin{tabular}{ccccccc}
\hline\hline
$|E_{^{19}{\rm C}}| $& $1/a$  & $b$  & $c$& $E_0$& $d$ &$r_{nc}$   \\
(keV) &  (fm$^{-1}$) &  (fm.keV)$^{-1}$ & (fm.keV$^2$)&  (keV)& (fm$^{-1}$) &(fm)   \\
\hline\hline
200 & -0.838 10$^{-2}$  & 5.861 10$^{-4}$  & 3.822 10$^{-8}$& 1304.0& 0.8376 &0.157\\
400 & -0.610 10$^{-1}$  & 6.719 10$^{-4}$  & 6.228 10$^{-8}$& 748.98& 0.5992 &0.174\\
600 & -2.491 10$^{-1}$  & 9.049 10$^{-4}$  & 1.276 10$^{-7}$& 402.94& 0.6344 &0.190\\
800 & -1.785            & 4.215 10$^{-3}$  & 8.114 10$^{-7}$& 78.86 & 2.1224 &0.205\\
830 & -5.104            & 1.153 10$^{-2}$  & 1.738 10$^{-6}$& 28.98 & 5.4396 &0.208\\
\hline\hline
\end{tabular}
\label{table2}
 \end{table}
 
 \begin{table}[thb!]
  \caption{Effective-range parameters, obtained by fitting Eq.~(\ref{kcotd}) to Fig.~\ref{figkcotze}, when considering different values of $|E_{^{19}{\rm C}}|$ (first column) with  high-range  Yamaguchi potentials.}
\begin{tabular}{ccccccc}
\hline\hline
$|E_{^{19}{\rm C}}| $& $1/a$  & $b$  & $c$& $E_0$& $d$ &$r_{nc}$   \\
(keV) &  (fm$^{-1}$) &  (fm.keV)$^{-1}$ & (fm.keV$^2$)&  (keV)& (fm$^{-1}$) &(fm)   \\
\hline\hline
200 & -0.317 10$^{-1}$  & 4.131 10$^{-4}$  & 1.181 10$^{-8}$& 881.91& 0.4052   &2.736\\
400 & -0.379 10$^{-1}$  & 5.042 10$^{-4}$  & 1.332 10$^{-8}$& 537.66& 0.3128   &3.233\\
600 & -1.125 10$^{-1}$  & 5.723 10$^{-4}$  & 5.344 10$^{-8}$& 324.85& 0.3040   &3.720\\
800 & -0.381           & 1.246 10$^{-3}$  & 4.753 10$^{-7}$& 132.92 & 0.5550   &4.255\\
850 & -0.671            & 1.975 10$^{-3}$  & 8.891 10$^{-7}$& 85.59 & 0.8465   &4.403\\
\hline\hline
\end{tabular}
\label{table3}
 \end{table}

\begin{figure}[tbh!]
\begin{center}
\includegraphics[scale=0.6]{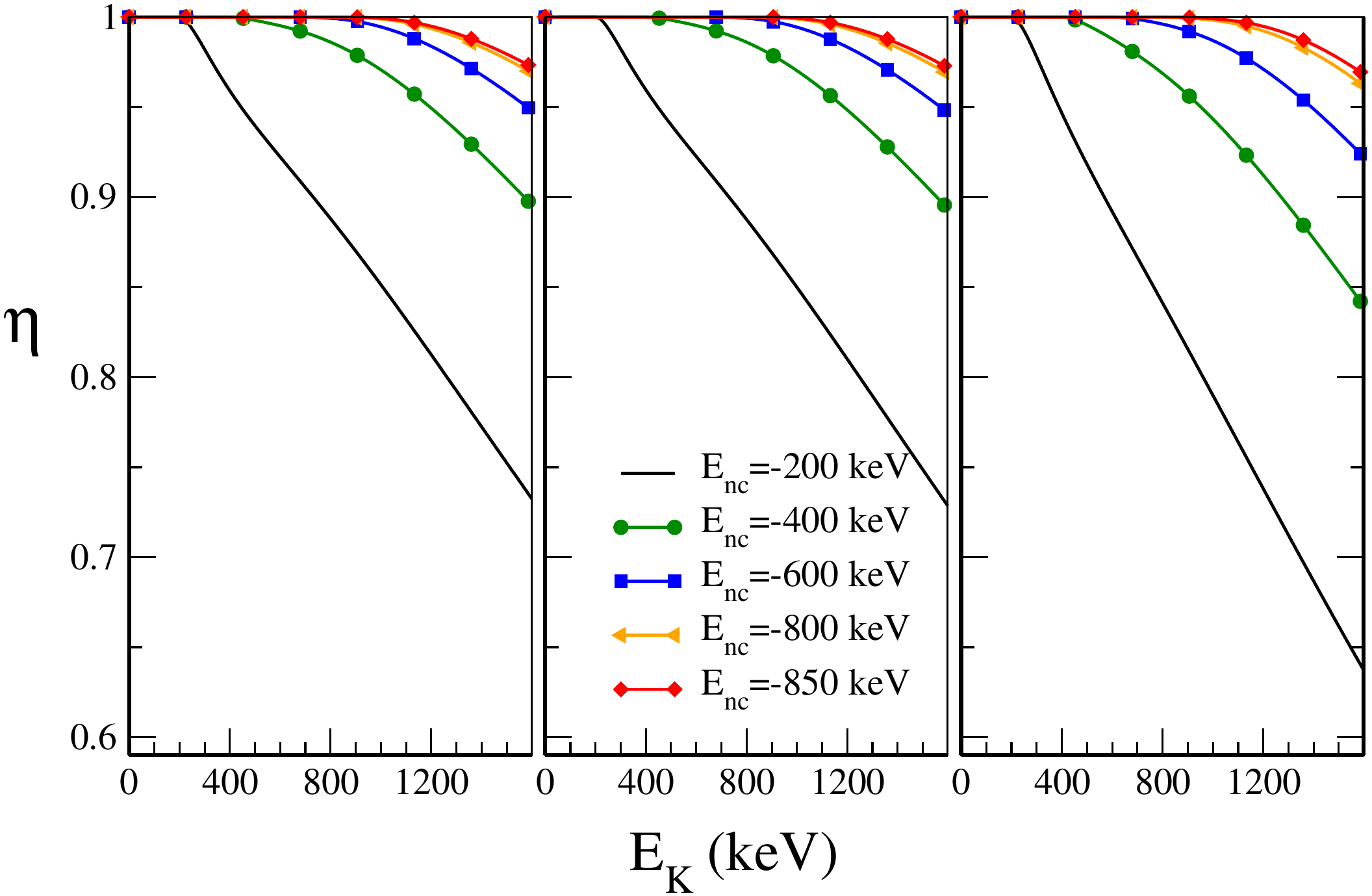}
\end{center}
\caption{ The $s-$wave absorption parameter $\eta$, shown in terms of the CM kinetic energy $E_K$ for zero-range potential (left frame), low-range Yamaguchi potential (middle frame, with large $\beta$ values), and high-range Yamaguchi potential (right frame, with small $\beta$ values.}
\label{figabs}
\end{figure}

In Fig.~\ref{figabs} we show that the absorption parameter, given by $\eta=|e^{2\i \delta_0}|= e^{-2\delta_0^I}$, is strongly 
affected by the binding energy of $^{19}$C. As observed, the absorption increases naturally with the size of $^{19}$C, as verified for the weakly-bound case with 200 keV.  By increasing the range in this figure
(decreasing the values of $\beta$ parameters), with the two-body interactions going from zero ranged (bottom panel), 
low ranged (middle panel with high $\beta$) and high ranged (top panel with low $\beta$), we also observe 
the increasing of the absorption parameter due to the size variation of $^{19}$C. In addition, for the same $^{19}$C energy the breakup is favored by increasing the interaction range, that seems reasonable as the long range attraction is weakened, making easier the transition to the inelastic channel, as given by the drop in the value of the inelasticity parameter, as shown in the figure.

\begin{figure}[tbh!]
\begin{center}
\includegraphics[scale=0.65]{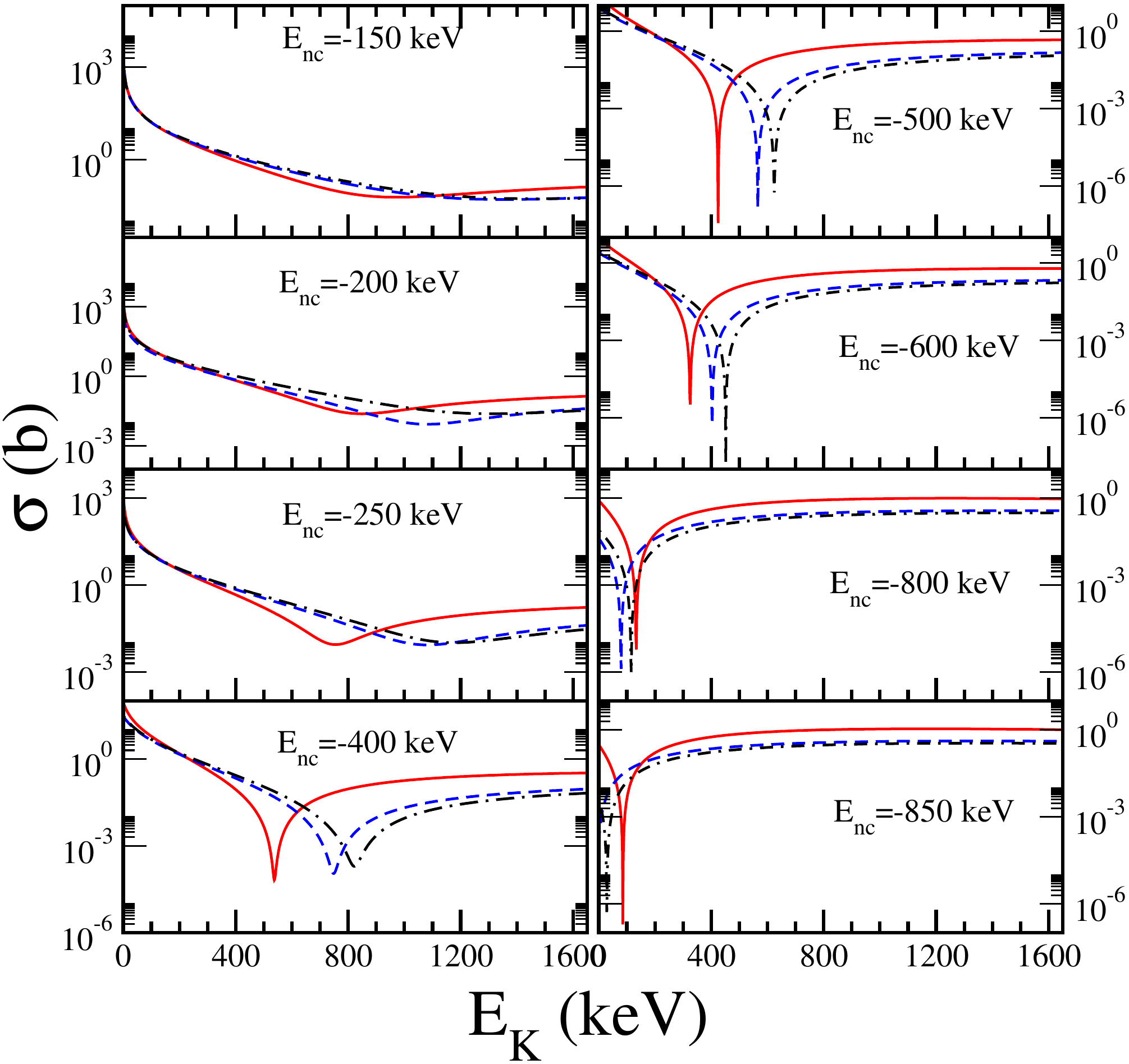}
\end{center}
\caption{ Evolution of the $n-\,^{19}$C elastic  cross-section for the $s-$wave scattering as a function of the center of mass kinetic energy
with the variation of $E_{^{19}{\rm C}}$ for fixed $E_{^{20}{\rm C}}$. The  results for Yamaguchi potentials are given by the solid lines (low $\beta$) and dashed-lines (high $\beta$). The dash-dotted lines give the results  for the zero-range model.
}
\label{cross}
\end{figure}

Our results for the $s-$wave scattering cross sections, obtained from $d\sigma/d\Omega$ 
$= \left| h_n(k;E)\right|^2$, are in Fig.~\ref{cross} (in logarithmic scales) as functions of $E_K$, by considering eight different values for $E_{nc}$, from -150 keV to -850 keV (as indicated inside the corresponding panels). 
The positions of the $k\cot\delta_0^R$ pole correspond to the observed minima  appearing  in the cross-section for  $E_K=E_0$, which is clearly seen when is $E_0$ is below the breakup threshold or when absorption to the breakup channel is not large. We observe that the minima will become more and more pronounced as the $^{19}$C binding increases, with the corresponding positions moving to smaller values of $E_K=E_0$, as illustrated in Fig. \ref{figpoloc19c20}.
However, in this trend (of moving the poles with the variation of $E_{^{19}{\rm C}}$) we should also notice that such behavior 
affects more the low-range cases. This reflects the slope changes already verified in Fig.~\ref{figpoloc19c20}, where one can
verify that $E_0$ moves faster with $E_{^{19}{\rm C}}$ for the zero-range results than for the finite-range cases. Therefore, at some 
point, for large enough $^{19}$C binding, the deep observed in the cross-section for the zero-range results will cross the high-range one.
The deep of the low-range potential is not crossing the zero-range  because the value of $E_0$ where this happens  is outside the physical 
scattering region. In addition, we observe a softening of the deep when $E_0>|E_{^{19}{\rm C}}|$ due flux absorption to the 
breakup channel, which is clearly observed in the figure for $^{19}$C binding energies below 400 keV. 

Except for the regions around the deeps, where we observe the strong variation of the cross-section, we note that its value are generally larger for the Yamaguchi interactions, and for the high-range  potential this trend is more visible. 
But apart such differences, the general behavior of these cross sections shows the same trend.

\section{Conclusions}\label{conclusions}

In the present work, we revisited the well-known study of $s-$wave phase-shift neutron-deuteron doublet 
scattering  in the context of the neutron-$^{19}$C scattering with a finite range potential in a neutron-neutron-core picture, extending a previous work done for a zero-range interaction model \cite{YamPLB08b}.
We restricted our study to $s-$wave observables, due to the fact that, when considering large scattering lengths and small 
effective ranges for a total zero-angular-momentum state, the properties of the discrete and continuum spectrum of 
the neutron-neutron-core system are dominated by the Efimov physics, namely the $s-$wave observables obeys scaling 
laws with dependence on the scattering lengths and the two-neutron separation energy of the halo system \cite{FrePPNP12}. 
Within this conceptual framework, we studied to which extent we can identify the general features brought by the 
universal scaling properties and Efimov physics in the neutron-$^{19}$C low-energy $s-$wave scattering when the two-body interaction ranges  are increased. For that, considering the three-body halo system $n-n-^{18}$C, we use a one term separable Yamaguchi potential having the physical value for the neutron-neutron effective range and 
for a neutron-core effective range close or larger than the core matter radius.

We investigated the low-energy properties of the elastic $s-$wave scattering for the neutron$-^{19}$C near the critical condition for the occurrence of an excited Efimov state. 
The results obtained in the present work are extending to finite-range interactions the ones obtained previously when considering the zero-range approach~\cite{YamPLB08b}. In view of
these results, confirming that the real part of the elastic $s-$wave phase shift ($\delta_0^R$) reveals a zero when the 
$n-n-c$ system is close to an excited Efimov state (bound or virtual) \cite{YamPLB08b}, we can also appreciate the effect of the range on the position of the zero. 

In close analogy with what happens to the $s-$wave phase-shift in the neutron-deuteron doublet scattering. It is well known that at 
low-energies neutron-deuteron doublet state is dominated by the Efimov physics, and the parameters of the effective range expansion of the neutron-deuteron doublet $s-$wave phase-shift present universal 
scaling laws with the triton binding energy for fixed nucleon-nucleon scattering lengths and effective ranges.

We establish to which extent the general universal scaling laws, strictly valid in the zero-range limit from the dominance of the Efimov physics, survive when finite range potentials are used to describe
 the neutron-$^{19}$C low-energy $s-$wave scattering. We studied the scaling of the real and the imaginary parts of the phase-shift with the $^{19}$C halo neutron binding energy when the two neutrons separation energy in $^{20}$C is fixed to its experimental value. 
The two quantities analyzed were $k\cot\delta^R_0$ ($\delta^R_0$ the real part of the phase-shift) and the inelasticity parameter $\eta$.  The first quantity exhibits a pole coming from the log periodic behavior of the scattering amplitude originated from the scale invariance of the three-body integral equations in the limit of a zero-range force 
and its breaking to a discrete scaling in the realm of the Efimov physics, 
which correlates the $s-$wave quantities to the binding energy of the three-body halo system, namely the $^{20}$C nucleus. 
That strong correlation survives for the finite range potential with effective neutron-neutron and neutron-core ranges compatible with physical values.  In the present context, the scaling law representing an observable depends on the binding energy of the three-body 
halo system, and n-n and n-core scattering lengths. Therefore, even with the $^{20}$C two-neutron separation energy and the neutron-neutron scattering length fixed, there is still dependence on the $^{19}$C binding energy, which was explored here having as a guideline the results for the zero-range model.

We verified that by considering a finite-range potential, the results for the $s-$wave scattering amplitude present  universal scaling features, with the variation of the $^{19}$C binding energy for fixed
$^{20}$C and neutron-neutron singlet virtual state energies. The scaling  of the effective range parameters and the
position of the pole of $k\cot{\delta_0^R}$ are in general accordance with previous results for the scaling obtained in the calculations with the zero-range potential. One striking feature is that the ratios
$\sqrt{E^B_{^{19}C}/E^0_{^{19}C}}$ obtained for finite range potentials from 0.44 to 0.45, where $E^0_{^{19}C}$ and $E^B_{^{19}C}$ being respectively, the values of the $^{19}$C energy where the pole is at the scattering threshold and the  excited state of $^{20}$C is at the scattering threshold, are quite close to $\approx 0.419 $ the universal ratio \cite{YamPLB08b}.  

The effect of the finite range appears by the softening of the corresponding variation in the positions of the pole, as well as the effective range parameters of neutron$-^{19}$C $s-$wave 
amplitude. The excited three-body $^{20}$C state turns into a virtual state for a large $^{19}$C binding, moving it from 167 keV to 190 keV when the effective ranges are increased to reasonable physical values. 
The increase of this quantity is physically consistent with the findings within Effective Field Theory \cite{hammer08}.
In addition, we have also clarified that the analytical structure 
of the unitary cut is not affected by the potential range or mass asymmetry of the three-body system.
Finally, we found that the deep in the elastic neutron-$^{19}$C $s-$wave cross-section comes close to the breakup threshold, when the halo-neutron separation energy in $^{19}$C is close to one of the experimental values of 530$\pm$130 keV from Ref. \cite{naka99}. The emergence of the universal features in the neutron-$^{19}$C $s-$wave 
elastic amplitude is evident when using potentials with physical effective ranges. 
The present study opens the prospect of investigating the transition amplitudes for three-body halo breakup to low-energy continuum states within the framework of scaling laws, in particular when monopole operators are involved in the transition.
Finally, in a perspective of actual experimental possibilities, along the lines as reviewed in Ref.~\cite{ulmanis}, we should also point out the 
relevance of an extension of the present investigation to heteronuclear ultracold quantum gases.
\section*{Acknowledgments}
This work was partly supported by funds provided by the Brazilian agencies Coordena\c c\~ao de Aperfei\c coamento de Pessoal
de N\'\i  vel Superior - CAPES [Proc. no. 88881.030363/2013-01(MTY and MAS) and a Senior Visitor Program at the Instituto Tecnol\'ogico de Aeron\'autica
(LT)], 
Conselho Nacional de Desenvolvimento Cient\'\i  fico e Tecnol\'ogico - CNPq [grants no. 302701/2013-3(MTY), 306191/2014-8(LT), 308486/2015-3 (TF)]
and Funda\c c\~ao de Amparo \'a Pesquisa do Estado de S\~ao Paulo - FAPESP [grant no. 2016/01816-2(MTY)].
 M.R.H. acknowledges the partial support by National Science Foundation under Contract No. NSF-HRD-1436702 with Central State University and by the Institute of Nuclear and Particle Physics at Ohio University.


\begin{thebibliography}{40}
\bibitem{FrePPNP12} 
T. Frederico, A. Delfino, L. Tomio, M.T. Yamashita, Prog. Part. Nucl. Phys. {\bf 67} (2012) 939.

\bibitem{Efimov} V. Efimov, Phys. Lett. B {\bf 33} (1970) 563.

\bibitem{riisager2013} K. Riisager,  Phys. Scr. {\bf T152} (2013) 014001.

\bibitem{YamNPA04}  M.T. Yamashita, L. Tomio and T. Frederico, Nucl. Phys. A {\bf 735} (2004) 40.

\bibitem{FedorovPRL94} D.V. Fedorov, A.S. Jensen and K. Riisager, Phys. Rev. Lett. {\bf 73} (1994) 2817.

\bibitem{Amorim97} A.E.A. Amorim, T. Frederico and L. Tomio, Phys. Rev. C {\bf 56} (1997) R2378.

\bibitem{hammer08} 
 D.L. Canham, H.-W. Hammer, Eur. Phys. J. A 37 (2008) 367; Nuclear Phys. A {\bf 836} (2010) 275.

\bibitem{MaccFBS15} A. O. Macchiavelli,  Few-Body Syst. {\bf 56} (2015) 773.
 
\bibitem{FRIB-MSU} Facility for Rare Isotopes Beams, Michigan State University, www.frib.msu.edu.

\bibitem{MaccLOI14} A.O. Macchiavelli, et al. NSCL/ReA3 letter of intent 14072 (2014).

 \bibitem{reiner} A.S. Reiner, Phys. Lett. B {\bf 28} (1969) 387.
\bibitem{wfuda} J.S. Whiting and M.G. Fuda, Phys. Rev. C {\bf 14} (1976) 18.
\bibitem{gfuda} B.A. Girard and M.G. Fuda, Phys. Rev. C {\bf 19} (1979) 579.
\bibitem{vanoers} W.T.H. van Oers and J.D. Seagrave, Phys. Lett. B {\bf 24} (1967) 562.
\bibitem{adhprc821} S.K. Adhikari, A.C. Fonseca, and L. Tomio, Phys. Rev. C {\bf 26} (1982) 77; 
S.K. Adhikari and L. Tomio, Phys. Rev. C {\bf 26} (1982) 83; S.K. Adhikari, L. Tomio and A.C. Fonseca, 
Phys. Rev. C {\bf 27} (1983) 1826. 
\bibitem{c20}  M.T. Yamashita, T. Frederico and L. Tomio, Phys. Rev. Lett. {\bf 99} (2007) 269201; 
Phys. Lett. B {\bf 660} (2008) 339.
\bibitem{braaten} E. Braaten and H.-W. Hammer, Phys. Rep. {\bf 428} (2006) 259.
\bibitem{YamPLB08b}  M.T. Yamashita, T. Frederico and L. Tomio, Phys. Lett. B {\bf 670} (2008) 49.
\bibitem{ArPRC04} V. Arora, I. Mazumdar, V.S. Bhasin, Phys. Rev. C {\bf 69} (2004) 061301(R); 
I. Mazumdar, A.R.P. Rau, V.S. Bhasin, Phys. Rev. Lett. {\bf 97} (2006) 062503.
\bibitem{jensen} M. Th\o gersen, D. V. Fedorov, A. S. Jensen, B. D. Esry and Yujun Wang,
Phys. Rev. A {\bf 80} (2009) 013608.

\bibitem{KieFBS11} A. Kievsky, E. Garrido, C. Romero-Redondo and P. Barletta, Few-Body Syst. {\bf 51} (2011) 259.

\bibitem{KiePRA15} A. Kievsky and M. Gattobigio, Phys. Rev. A {\bf 92} (2015) 062715.

\bibitem{KiePRA16}  R. \'Alvarez-Rodr\'\i guez, A. Deltuva, M. Gattobigio and A. Kievsky, Phys. Rev. A
{\bf 93} (2016) 062701.

\bibitem{audi}
G. Audi, A.H. Wapstra, and C. Thibault, Nucl. Phys. A {\bf 729} (2003) 337.
\bibitem{naka99} T. Nakamura et al., Phys. Rev. Lett. {\bf 83} (1999) 1112.
\bibitem{OzaNPA01} A. Ozawa et al., Nucl. Phys. A {\bf 691}, 599 (2001).
\bibitem{adhprc81} L. Tomio and S.K. Adhikari, Phys. Rev. C {\bf 22} (1980) 28; 
{\bf 22} (1980) 2359; {\bf 24} (1981) 43; S.K. Adhikari and L. Tomio, 
Phys. Rev. C {\bf 24} (1981) 1186.
\bibitem{MilPRP90} G. A. Miller, M. K. Nefkens, and I. Slaus, Phys. Rep. {\bf 194} (1990) 1.
\bibitem{ulmanis} J. Ulmanis, S. H\"afner, E. D. Kuhnle and M. Weidem\"uller, Nat. Sc. Rev. {\bf 3} (2016) 174.
\end{thebibliography}
\end{document}